\documentclass[useAMS,usenatbib]{mn2e}

\usepackage[dvips]{graphicx}
\usepackage{graphics}
\usepackage{psfig}
\usepackage{lscape}
\usepackage{amssymb}
\usepackage{mathrsfs}
\input{epsf}

\title[X-ray properties of star forming galaxies at $z\sim$1]
{X-ray properties of UV-selected star forming galaxies at $z\sim$~1
in the Hubble Deep Field North}
\author[E. S. Laird et al.]
{E. S. Laird,$^{1}$\thanks{E-mail:~e.laird@imperial.ac.uk;}
K. Nandra,$^{1}$ K. L. Adelberger,$^{2}$ C. C. Steidel,$^{3}$ N. A. Reddy$^{3}$\\ 
$^{1}$Astrophysics Group, Imperial College London, Blackett Laboratory, Prince Consort Road, London SW7 2AZ, UK\\
$^{2}$Carnegie Observatories, 813 Santa Barbara Street, Pasadena, CA 91101, USA\\
$^{3}$California Institute of Technolgy, MS 105-24, Pasadena, CA 91125, USA\\}
                               
\begin{document}

\date{Accepted January 2005.}
 
\pagerange{\pageref{firstpage}--\pageref{lastpage}} \pubyear{2004}

\maketitle

\label{firstpage}
\begin{abstract} We present an analysis of the X-ray emission from a
large sample of ultraviolet (UV) selected, star forming galaxies with
$0.74<z<1.32$ in the \textit{Hubble Deep Field} North (HDF-N)
region. By excluding all sources with significant detected X-ray
emission in the 2~Ms \textit{Chandra} observation we are able to
examine the properties of galaxies for which the dominant emission in
both UV and X-ray is expected to be predominantly due to star formation. 
Stacking the X-ray flux from 216 galaxies in the soft and  
hard bands produces significant detections (14.9$\sigma$ and
3.2$\sigma$, respectively). 
The derived mean 2--10~keV rest-frame luminosity is
$2.97\pm0.26\times10^{40}$~erg~s$^{-1}$, corresponding to an X-ray
derived star formation rate (SFR) of $6.0\pm0.6$ M$_{\sun}~\rmn{yr}^{-1}$.  
Comparing the X-ray value with the mean UV derived SFR, uncorrected for attenuation,
we find that the average UV attenuation correction factor is $\sim$3. 
By binning the galaxy sample according to UV magnitude and colour, and stacking the
observed frame soft band X-ray flux in each bin, correlations between UV and X-ray
emission are examined.
We find a strong positive correlation between X-ray emission and rest-frame UV
emission, consistent with a strict linear relationship,
L$_{\rmn{X}}\propto\rmn{L}_{\rmn{UV}}$, at the 90 per cent level.
A correlation between the ratio of X-ray-to-UV emission and UV colour
is also seen, such 
that  $\rmn{L}_{\rmn{X}}/\rmn{L}_{\rmn{UV}}$ increases for redder galaxies. 
We find no direct relation between X-ray flux and UV colour. 
Given that X-ray emission offers a view of star formation regions that
is relatively unaffected by extinction, results such as these can be
used to evaluate the effects of dust on the UV emission from high-$z$
galaxies. For instance, using the observed correlation between UV
colour excess and the ratio of X-ray-to-UV emission -- a measure of UV
obscuration -- we derive a relationship for estimating UV attenuation
corrections as a function of colour excess.
The observed relation is inconsistent with the \citet{cal00}
reddening law which over predicts the range in UV attenuation
corrections by a factor of $\sim$100 for the UV selected $z$$\sim$1
galaxies in this sample. 
\end{abstract}


\section{Introduction}

The vast majority of known high redshift galaxies are selected in the
rest-frame ultraviolet (UV). For instance the Lyman break technique,
which selects actively star forming galaxies based on their broad band
ultraviolet (UV) colours, is
the primary method for efficiently selecting redshift $z$$\gtrsim$3
galaxies (e.g. \citealt{steidel96}; \citealt{steidel99};
\citealt{steidel03}) and has inspired similar techniques to isolate
star forming galaxies from $1\lesssim z \lesssim3$ (\citealt{adel04};
\citealt{steidel04}).  The galaxies identified in these surveys are
observed  during the peak of the cosmic star formation rate density
(e.g  \citealt{lilly96}; \citealt{madau96}; \citealt{hughes98}) and
represent a significant fraction of the total star formation at high
redshift \citep{adel00}. The star formation rates (SFRs) derived
from these UV-selected galaxies are therefore crucial for determining
the history of star formation and heavy element production in the
Universe.  

However the SFRs obtained from the rest-frame UV fluxes  are
very sensitive to the effects of dust attenuation intrinsic to the
galaxies \citep{kennicutt98}. Correcting the observed fluxes for the
attenuation requires an understanding about the true
nature of the sources and their dust content that is currently
lacking. To overcome this, average 
corrections are applied based on dust attenuation laws observed in
local starburst galaxies (e.g. \citealt*{cal94}; \citealt{cal00};
\citealt*{meurer99}, hereafter MHC99), which appear to be the 
local analogues of the high redshift LBGs (e.g. \citealt{steidel96};
\citealt{giavalisco02}). However there is considerable disagreement
about the size of the corrections to be applied with some evidence that
attenuation has been overestimated \citep{bell02}.

X-ray observations can add to this debate by providing a view of
galaxies that is relatively unaffected by dust and extinction.
Hard X-rays are emitted from star forming galaxies by 
a combination of  high mass X-ray binaries (HMXBs), supernovae
remnants and hot gas \citep{fabbiano89}. Locally strong
correlations between X-ray luminosity and other SFR measures have been
seen \citep*{david92}, showing that X-ray luminosity can be used as a SFR measure
(\citealt*{ranalli03}; \citealt*{grimm03}; \citealt{N02}). By
comparing X-ray derived 
SFRs with UV estimates we can independently check the validity of the
attenuation corrections and gain insight into the nature and dust
content of the UV-selected galaxies in the early Universe.

Although the majority of high redshift star forming galaxies are not
bright enough to be detected with the current class of X-ray
telescopes, with deep \textit{Chandra} observations the average
properties of a sample can be determined using stacking techniques
(e.g. \citealt{horn01}; \citealt{brandt01};
\citealt{N02}). Several efforts at this have already been made.
\citet{brandt01} stacked the emission from 24 $z$$\sim$3 LBGs in the 
1~Ms Chandra Deep Field-North (CDF-N), detecting a significant
0.5--2~keV observed-frame signal. Using local starburst reddening
relations \citet*{seibert02} predicted the mean X-ray luminosity of the
24 LBGs and found it agreed well with the stacking results. Also in
the 1~Ms CDF-N, \citet{N02} analysed the emission from two much larger
samples of 148 $z$$\sim$3 LBGs and 95 $z$$\sim$1 star forming
galaxies. More recently using the 2~Ms data in the GOODS-North field,
\citet{reddy04} analysed the stacked emission from a sample of
UV-selected $z$$\sim$2 galaxies and
\citet{lehmer04} analysed the emission from large samples of
$z$$\simeq$3, 4, 5 and 6 LBGs. 

In this paper we seek to add to this work by improving upon and
extending the analysis of the $z$$\sim$1 galaxies by Nandra et
al. (2002; hereafter N02) using the 2~Ms observation of the HDF-N
region.  
Using a much larger catalogue of 255 $z$$\sim$1 UV-selected Balmer
break galaxies (BBGs) we assess the X-ray properties and the relations
between UV and X-ray emission of the galaxies, during a period 
where the star formation density of the Universe was at its peak.
Throughout, we assume a standard, 
flat $\Lambda$CDM cosmology with $\Omega_{\Lambda}=0.7$ and
H$_{0}=70$ km s$^{-1}$ Mpc$^{-1}$.

\section{Data and analysis}

\subsection{Optical data}
The BBG candidates were colour selected from optical images
using photometric criteria, detailed in \citet{adel04}.
Briefly, the BBGs have been selected based on the strengths of the
Balmer and 4000~\AA~breaks in their spectral energy distributions,
designed to identify 
starburst, Sb- and Sc-type star forming galaxies \citep{adel04}. The
candidate list, which includes all BBGs with spectroscopically
confirmed redshifts with $0.74<z<1.32$, is considerably larger than that
used in N02 thanks to the extensive spectroscopic effort
of the  Team Keck Treasury Redshift Survey in the Great Observatories
Origins Deep Survey (GOODS)-North field   
\citep{wirth04}. There are 255 galaxies in our CDF-N BBG sample, with
$\overline z$=0.95. $U_n,G,\mathscr{R}$ and $I$ magnitudes and 
colours have been measured for all the objects.

\subsection{X-ray data}
The HDF-N region has been observed with \textit{Chandra} ACIS-I 20
times since launch, with the total exposure time approximately 2~Ms. 
Analysis of these observations, and a  point source
catalogue, was presented in Alexander et al. (2003; hereafter A03). 
For this paper we used the raw data available from the public archive.

The HDF-N was observed over a period of 27 months during which
time the observing conditions and operating mode of the telescope
changed. The 20 observations can broadly be separated into three
groups: the first three observations were taken in Faint mode with a
focal plane temperature of $-110$\degr C, the following 9 observations 
also in Faint mode with temperature $-120$\degr C, and the final 8
observations, constituting the second 1~Ms, taken in Very Faint mode
at -120\degr~C. The differences in the observing modes  at times
warranted  separate treatment when reducing the data.

Our analysis is restricted to an approximately
$10\farcm3\times10\farcm3$ region within the larger ACIS-I field of
view. This encompasses the optical BBG survey region
($8\farcm7\times8\farcm7$). 

\subsection{Data reduction}

Data reduction was carried out using the \textit{Chandra} X-ray Center
(CXC) \textit{Chandra} Interactive Analysis of Observations ({\small
CIAO}) software, version 3.0.1, and the \textit{Chandra} calibration
database (CALDB) version 2.23.  Using the tool {\small
ACIS\_PROCESS\_EVENTS}, all observations had up-to-date gain maps and
geometry solutions applied. In addition, the observations taken at
-120\degr~C were corrected for radiation damage inflicted in the first
few months of \textit{Chandra} operations using the standard CXC CTI
correction, and the eight observations taken in Very Faint mode were
further cleaned by identifying likely background events.
The event files were filtered using standard
screening criteria -- ASCA event grades 0,2,3,4,6, Good Time Intervals
and status=0 -- to remove bad and flaring pixels, bad columns,
cosmic ray afterglows etc.
Source subtracted lightcurves were
created and periods where the background was more than 3$\sigma$ from
the mean value were flagged
using {\small ANALYZE\_LTCRV}. 
All the periods of high background
identified by  {\small ANALYZE\_LTCRV} were removed 
in addition to a further 44.4~ks from observation 2344 where
the procedure failed to identify a flare  and 
15.2~ks from observation 3389 where it failed to
identify the full duration of a flare.

Before coadding the event files, the relative astrometry was improved
by registering each observation to the coordinate frame of observation 
1671 using bright sources within 6 arcminutes of the aim points. This
was done using the {\small ALIGN\_EVT} tool written 
by T. Aldcroft\footnote{See
http://cxc.harvard.edu/cal/ASPECT/align\_evt/ for details.}.
The root-mean-square shift applied to the observations was 0.54
arcseconds in RA 
and 0.38 arcseconds in Dec. 
The average aim point for the combined
observations, weighted by exposure time,  was 
$\alpha=12^{\rm h}36^{\rm m}47\fs59$, 
$\delta=+62\degr14\arcmin08\farcs1$ (J2000.0) and the total exposure
time was 1.86~Ms.  

\begin{figure}
\begin{center}
\includegraphics[width=75mm]{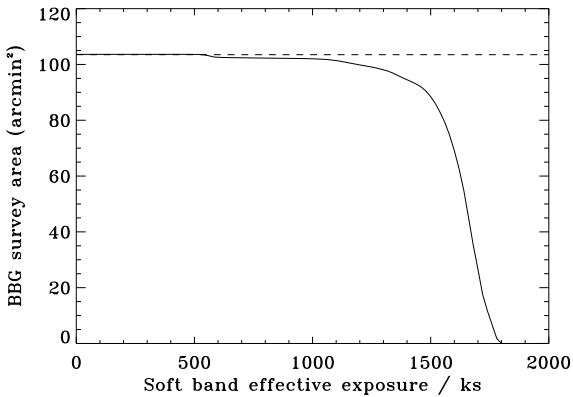}
\caption{The BBG survey area versus the minimum effective
exposure in the soft-band exposure map. The dashed line shows the
solid angle of the region of our analysis. Over 98 per cent of the BBG area
has at least 1~Ms effective exposure and over 85 per cent has at least
1.5~Ms.}
\label{effexp}
\end{center}
\end{figure}

We constructed event files and images in four bands: 0.5--2~keV (soft
band),  0.5--7~keV (full band), 2--7~keV (hard band) and 4--7~keV
(ultra-hard band). Effective exposure maps were constructed
using the {\small MERGE\_ALL} tool for each of the four bands -- these
take into account the effects of 
vignetting, gaps between chips, bad column and bad pixel
filtering and are required for converting observed counts to flux.
 The maps were created at a single energy representative of
the photons in each band: 1~keV for the soft, 2.5 keV for the full,
4~keV for the hard and 5.5~keV for the ultra-hard band. These are the
mean photon energies of detected sources in the band. Given that bad
pixel files for observations in different modes vary considerably, and
that {\small MERGE\_ALL} will only accept one bad pixel file, in order
to properly account for their effect we
created separate exposure maps for each of the three groups of observations
described  above. These maps were then  re-gridded  to produce effective exposure
maps for the full $\sim2$Ms.  A plot of solid angle over the
BBG survey area  as a
function of soft band effective exposure is shown in
Figure 1.

When performing photometry with \textit{Chandra} images the
energy-dependent and position-dependent point-spread function (PSF)
must be properly accounted for. Due to the different aim points of the
20 observations in the HDF-N a given sky position will have a
different PSF in each observation and the final PSF should therefore
represent the average value. To calculate the PSF for a given sky
position we used the {\small MKPSF} tool with the CALDB PSF library
file {\small acisi1998-11-052dpsf1N0002.fits} to create an image of
the PSF at the corresponding detector position in each of the
observations and then combined the images, weighted by exposure, to
create an average PSF image. A radial profile of the PSF image was
then obtained by extracting counts from 30 concentric circles out to a
maximum radius of 30 arcseconds. Assuming all counts are contained
within the 30 arcsecond radius we then calculated by interpolation the
radius at which, say, 90 per cent of the counts are enclosed (referred
to as the 90 per cent encircled energy fraction, EEF) and also the
fraction of counts enclosed within a given fixed radius, say, 1.5
arcseconds\footnote{The assumption that all counts will be contained
within 30 arcseconds may not always be strictly valid, particularly at
large off axis angles (OAAs) and at high energies. However at the
moderate OAAs considered in this work ($\simeq6$ arcminutes) any error
introduced is likely to be very minor.}. This method can be repeated
for different energies, positions, EEFs and radii.

Finally, PSF-corrected count rates should be converted to flux. For
this conversion we have 
assumed a power-law source spectrum with $\Gamma$=2 in the 
all bands  and
Galactic column density N$_{\rm H}=1.6\times10^{20}$~cm$^{-2}$ \citep{stark92}. 
When converting from counts to flux the 
full, hard and ultra-hard bands were extrapolated to the standard
upper-limit of 10~keV. 

It has been found that the \textit{Chandra} ACIS quantum efficiency
(QE) has been suffering continuous degradation since launch -- most
likely a result of the build-up of material on the ACIS detectors or
optical blocking filers \citep{marshall04} -- which affects the spectral response of the
detectors and, therefore, the counts to flux conversion. 
We have calculated the necessary corrections
to the fluxes using the {\small ACISABS} model, version
1.1-2\footnote{For more information on the ACIS QE degradation see
http://cxc.harvard.edu/cal/Acis/Cal\_prods/qeDeg/.}$^{,}$\footnote{To 
take into account the time dependence of the degradation
we applied the {\small ACISABS} correction to the response files for
each individual observation then combined to obtain the 
exposure-weighted average correction.}.  For Galactic column density
and $\Gamma$=2 the soft, hard, full and ultra-hard fluxes should be
increased by 14.7, 0.6, 11.5 and 0.1 per cent respectively. All fluxes
quoted in this paper have been corrected for the QE degradation.

\subsection{Source detection}
Source detection was carried out in the field, the reasons for which
are threefold: 
(1) identify BBG galaxies that are directly detected at 2~Ms, 
(2) use these objects to determine any overall offsets between the
coordinate reference frames of the optical and X-ray 
images\footnote{In order to increase signal-to-noise ratio any offset
between coordinate reference frames should 
be corrected for before performing stacking analysis.}, 
and 
(3) identify all X-ray sources for exclusion from the stacking
background calculations.
Source detection was performed using our own procedure, which is
detailed in \citealt{nandra05}.
To identify X-ray sources for exclusion from
the stacking background determination we set the false
probability threshold to $10^{-6}$, a level for which 1 spurious
source is expected per band in the $10\farcm3\times10\farcm3$ survey area.
To find only X-ray counterparts to the BBGs the
threshold can be set much lower without introducing likely spurious
X-ray sources as the area in question in considerably reduced. 
We used a threshold of $2\times10^{-4}$ which, given the number of
BBGs and the area over which the BBG and \textit{Chandra} catalogues are
cross-correlated, results in a probability of a false X-ray detection
of a BBG of less than 0.15 per cent per detection band. In each case
sources were detected in the full, soft, hard and ultra-hard bands and
a band merged catalogue produced.

The lower significance band merged \textit{Chandra} catalogue was matched to the BBG
catalogue to search for any X-ray counterparts and to identify
any overall shift in the optical and X-ray reference frames that
should be removed before performing stacking. A small overall offset
between the two catalogues was found and the positions of the BBG
candidates were shifted by -0.47 arcseconds in RA and -0.88 arcseconds
in Dec to agree with the reference frame of the \textit{Chandra}
observation. 

After correcting for this overall offset we re-matched the
band merged \textit{Chandra} catalogue to the BBG catalogue,
using a radius of 1.5 arcseconds, to search for X-ray counterparts.

\subsection{Stacking procedure}

To determine the mean X-ray properties of the Balmer break galaxies
that are too weak to be directly detected we employ a stacking
technique similar to those described in N02 and \citet{brandt01}. 

\begin{figure}
\begin{center}
\includegraphics[width=75mm]{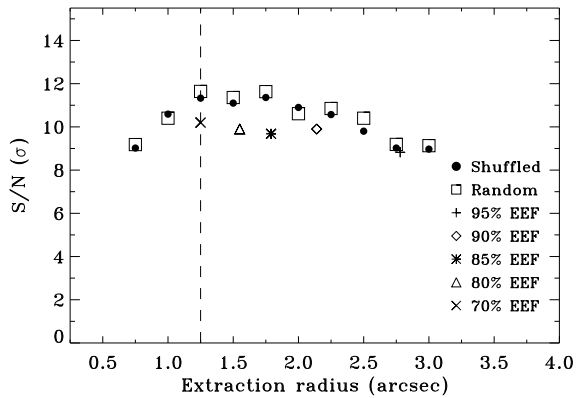}
\caption{ 
The signal-to-noise ratio vs. extraction radius in arcseconds. The
signal-to-noise ratio is an inverse measure of the fractional error on
the flux and is given by 
$({\rmn{S/{\sqrt{S+B}}}})$, where S and B
are the net source counts and background counts respectively. 
The two
different background methods, shuffled and random positions, give very
similar results. The results using a fixed fraction of the EEF are
plotted at the average extraction radius used.  The vertical dashed
lines denote our chosen extraction radius of 1.25 arcseconds.
} 
\label{extr_rad}
\end{center}
\end{figure}

We first identify all the candidates that
have an X-ray counterpart and exclude them from the stacking.
Secondly we identify any candidates with
an unassociated, nearby X-ray source -- 
these objects will be affected by the
low-level PSF wings of the \textit{Chandra} PSF and will lead to
erroneously high source counts. We reject from the
stacking all candidates which are separated from an X-ray source by
a distance less than that of the 95 per cent EEF radius in that
region. In a given stacking sample normally zero or one and never
more than two sources were rejected. 
For the remaining galaxies in the samples we extract 
counts from the optical positions, summing both the detected counts
and the number of pixels used for the extraction.

The background counts contributing to the total source signal are
estimated in two ways. First we randomly shuffled the optical galaxy
positions by 5--10 arcseconds and extracted counts from an X-ray
source-masked image. Secondly we used random positions from anywhere
over the $10\farcm3\times10\farcm3$ region of interest and again
extracted counts from the masked image. In both cases,
this was repeated 1000 times for each galaxy -- sufficient to get an
accurate estimate of the background counts and the dispersion. The
counts and pixels were summed then scaled to the same area as the
source extraction. Both background procedures produced very similar
results -- in this paper we quote all results for the shuffled
positions, which should better account for any local variations in the
background level. 

The strength and accuracy of the observed signal is sensitive to the
size of extraction radius used. A highly significant soft band detection of a
smaller sample of BBGs in this field has already been achieved by N02: with
the deeper data we therefore seek an extraction radius 
to minimize the flux errors rather than maximise the significance 
of the detection.
We adopted an empirical approach to determining
the optimal extraction radius -- we tested 10 fixed extraction radii
between 0.75 and 3.0 arcseconds, as well as the 70, 80, 85, 90 and
95 per cent EEF radii (Figure 2), selecting the option yielding
the maximum signal-to-noise ratio. 
We chose a fixed extraction radius of 1.25 
arcseconds\footnote{We note that N02 contains an
error regarding the optimal extraction radius used. 
The paper states that the optimal radius found
was 2.5 arcseconds -- in fact 2.5 arcseconds was the optimal
extraction diameter.}, which gave a 14.9$\sigma$ detection with a
signal-to-noise ratio of 11.6 (we define detection significance as 
${\rmn{ S/{\sqrt{B}}}}$ and signal-to-noise ratio as $[{\rmn{
S/{\sqrt{S+B}}}}]$, where S and B are the net source and background
counts respectively).
The appropriate PSF corrections for a 1.25 arcseconds extraction
radius were calculated and the flux estimates
corrected accordingly. We used 
the full sample of BBGs for the stacking and did not limit the
analysis to sources at small off axis angles.

\section{Results}

\subsection{Directly detected sources} 

Of the 255 BBGs in our sample 37 are directly detected in X-ray. Some
of these sources show clear evidence of harbouring an AGN (i.e. broad
optical emission lines, X-ray luminosity greater than
$10^{43}~\rmn{erg~s^{-1}}$)
while for others the X-ray emission could be due to 
an AGN, star formation, or a combination of both. In this paper we are
concerned with the X-ray properties of galaxies where the emission is
likely to be dominated by normal star forming processes and we consequently seek to
omit from the analysis all galaxies where AGN emission may
dominate. We therefore, conservatively, exclude all 37
of the directly detected galaxies and deal only with the undetected
BBGs for the remainder of this paper.  
At $z$$\sim$1 this equates to removing all galaxies with  L$_{\rmn{2-10~keV}} \gtrsim
10^{41}~\rmn{erg~s^{-1}}$.
Full analysis of the X-ray, optical and UV properties of the detected
galaxies including starburst and AGN separation and identification will be
addressed in an upcoming paper (Laird et al., in preparation).

\subsection{Stacking results for undetected sources}

\begin{table*}
\begin{minipage}{176mm}
\caption{Stacking results of undetected BBGs  
for entire sample and sub-samples based on rest-frame
1800~\AA~emission. Col.(1): Galaxy sample. 
Col.(2): Observed-frame energy band. 
Col.(3): Number of galaxies included in stacking sample, taking into account rejected galaxies as described in \S2.5. 
Col.(4): Mean $\mathscr{R}$ magnitude. 
Col.(5): Mean redshift. 
Col.(6): Signal-to-noise ratio $[{\rmn{ S/{\sqrt{S+B}}}}]$, where S
and B are the net source and background counts respectively.
Col.(7): X-ray flux per galaxy; 0.5--2~keV for soft band, 2--10~keV for hard band.
Col.(8): X-ray luminosity per galaxy in the 2--10~keV band, derived
from soft band flux assuming $\Gamma$=2.0 and Galactic
N$_{\rmn{H}}$. 10--50~keV luminosity is given for the hard band, also assuming $\Gamma$=2.0.
Col.(9): SFR from 2--10~keV luminosity~\citep{ranalli03}. Errors are statistical only.
Col. (10): Ratio of X-ray derived SFR to UV SFR, uncorrected for attenuation.
}
\label{stacking}
\begin{tabular}{@{}llcccrcccc@{}}
\hline
Sample & Band & N & $\langle \mathscr{R} \rangle$ & $\langle z \rangle$ & S/N & F$_{\rmn{X}}$ & L$_{\rmn{X}}$ & $\langle SFR \rangle$ & SFR$_{\rmn{X}}$/SFR$_{\rmn{UV}}^{\rmn{uncor}  }$\\
       &      &   &                 &       &     & ($10^{-18}~\rmn{erg~cm^{-2}~s^{-1}}$) & ($10^{40}~\rmn{erg~s^{-1}}$) & (M$_{\sun}~\rmn{yr}^{-1}$) & \\
(1) & (2)  & (3)  & (4)  & (5)  & (6)  & (7)  & (8)  & (9) & (10)\\
\hline
Undetected BBGs   & Soft & 216 & 23.42 & 0.95 & 11.2 &$5.51\pm0.49$  & $2.97\pm0.26$ & $6.0\pm0.6$ & $3.0\pm0.4$\\
Undetected BBGs   & Hard & 216 & 23.42 & 0.95 & 3.0  &$8.87\pm2.91$ & $7.07\pm2.32$       & \ldots& \ldots \\
22.57 $<U_n\le$ 23.72 & Soft & 34  & 22.50 & 0.91 & 6.7  &$10.21\pm1.53$ & $4.81\pm0.72$ & $9.6\pm1.4$ & $1.8\pm0.3$\\
23.72 $<U_n\le$ 24.13 & Soft & 30  & 22.96 & 0.92 & 5.7  &$ 8.81\pm1.44$ & $3.83\pm0.68$ & $7.7\pm1.4$ & $2.6\pm0.5$\\
24.13 $<U_n\le$ 24.40 & Soft & 38  & 23.35 & 0.97 & 5.2  &$ 6.25\pm1.21$ & $3.37\pm0.65$ & $6.7\pm1.3$ & $2.7\pm0.5$\\
24.40 $<U_n\le$ 24.71 & Soft & 40  & 23.63 & 0.97 & 3.1  &$ 3.25\pm1.04$ & $1.75\pm0.56$ & $3.5\pm1.1$ & $1.8\pm0.6$\\
24.71 $<U_n\le$ 25.25 & Soft & 39  & 23.83 & 0.96 & 3.9  &$ 4.02\pm1.03$ & $2.16\pm0.56$ & $4.3\pm1.1$ & $3.2\pm0.8$\\
25.25 $<U_n\le$ 27.03 & Soft & 35  & 24.16 & 0.97 & 2.5  &$ 2.63\pm1.07$ & $1.42\pm0.58$ & $2.8\pm1.2$ & $4.4\pm1.8$\\
\hline
\end{tabular}
\end{minipage}
\end{table*}

\begin{figure}
\begin{center}
\includegraphics[width=75mm]{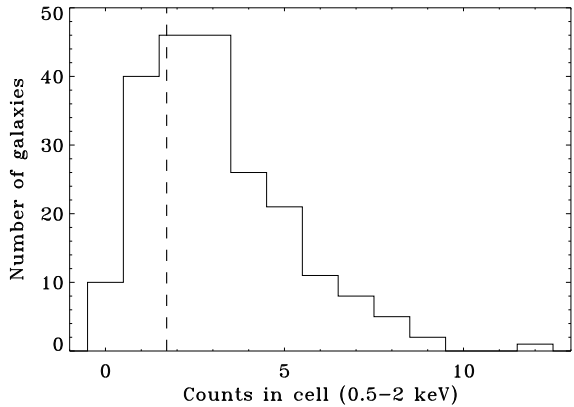}
\caption{Count distribution for the undetected BBGs in the soft
band. The vertical line denotes the mean background counts per count
extraction cell, derived via the shuffle background method.  }
\end{center}
\label{counts}
\end{figure}

Stacking the emission from the undetected BBGs produces
significant detections in both the soft and hard bands (Table 1).
In the soft band, stacking the counts from 216 undetected BBGs\footnote{Two 
undetected BBGs are rejected from the stacking as described in \S2.6}   
produces a total 
of 290.8 net counts and a detection significance of 14.9.
An average of 1.4 net counts per galaxy are detected, with 1.76
mean background counts in each galaxy extraction cell.
The distribution of total counts (S+B) in an extraction cell
(Figure 3) covers a wide range, with an almost continuous
distribution of counts from the peak at 2--3 counts per cell up to
$\sim$12 counts per cell and the realm of direct detections. 
Figure 3 demonstrates the stacking signal is not
dominated by a few bright sources but instead well represents the
entire sample.
The detected soft band signal for the BBGs corresponds to
a mean flux per galaxy of
$5.51\pm0.49\times10^{-18}$~erg~s$^{-1}$~cm$^{-2}$ (0.5--2~keV) and a
mean luminosity of 
L$_{2-10 \rmn{keV}} = 2.97\pm0.26\times10^{40}$~erg~s$^{-1}$.
The mean flux per undetected BBG found here
is in agreement with that found with the 1~Ms \textit{Chandra} data
for the smaller sample of BBGs by N02. 

A low significance 3.2$\sigma$ signal is detected from stacking the hard band
emission from 216 undetected BBGs. There are 87.4 total net hard band
counts detected, corresponding to an average of 0.40 net counts per
galaxy. The expected background level is 3.4 mean counts
per extraction cell.  This weak but significant signal results in a
mean 2--10~keV flux per BBG of 
$8.87\pm2.91\times10^{-18}$~erg~s$^{-1}$~cm$^{-2}$. 
The spectral constraints implied by this detection are 
discussed in \S4.1

\subsection{X-ray and UV correlations for stacked BBGs}

\begin{table*}
\begin{minipage}{125mm}
\begin{center}
\caption{Spearman's rank and linear Pearson's correlation tests.
Col.(1): The two quantities being tested. Cols. (2) and (3): Spearman rank
correlation coefficient and chance probability. 
Cols. (4) and (5): Linear Pearson
correlation coefficient and chance probability. 
In each test a coefficient of -1, 0 and +1 indicates a
negative correlation, no correlation and a positive correlation,
respectively.   
}
\label{coeff}
\begin{tabular}{@{}lcccc@{}}
\hline
Test pair & Spearman rank  & & Linear Pearson & \\
         & Coefficient & Prob. & Coefficient & Prob. \\
(1) & (2) &(3) &(4) &(5) \\
\hline
log f$_{\rmn{x}},~$U$_{\rmn{n}}$ & -0.943 & $<$0.005 & -0.927 & 0.008\\
log f$_{\rmn{x}},~$G           & -0.886 & 0.019    & -0.934 & 0.006\\
log f$_{\rmn{x}},~$$\mathscr{R}$ & -0.943 & $<$0.005 & -0.953 & $<$0.005\\
log f$_{\rmn{x}},~$I           & -1.000 & $<$0.005 & -0.988 & $<$0.005\\
log f$_{\rmn{x}},~$($U_{\rmn{n}}-G$) & -0.257 & 0.623  &  0.143 & 0.787\\
log f$_{\rmn{x}},~$($G-\mathscr{R}$) &  0.428 & 0.397  &  0.468 & 0.349 \\
log f$_{\rmn{x}},~$$(\mathscr{R}-I$) &  0.114 & 0.623  &  0.114 & 0.829 \\
log L$_{\rmn{x}},~$log L$_{\rmn{UV}}$ & 0.943  & $<$0.005 &  0.915   & 0.010 \\
log(SFR$_{\rmn{x}}/\rmn{SFR}_{\rmn{UV}}),~$E(1800\AA--2400\AA) & 0.486  & 0.329   &  0.855   & 0.030 \\
log(SFR$_{\rmn{x}}/\rmn{SFR}_{\rmn{UV}}),~$E(2400\AA--3400\AA) & 0.771  & 0.072   &  0.892   & 0.017 \\
log(SFR$_{\rmn{x}}/\rmn{SFR}_{\rmn{UV}}),~$E(3400\AA--4050\AA) & 0.714  & 0.111   &  0.840   & 0.036 \\
\hline
\end{tabular}
\end{center}
\end{minipage}
\end{table*}

\begin{figure*}
\begin{center}
\includegraphics[width=12.0cm, height=8.3cm]{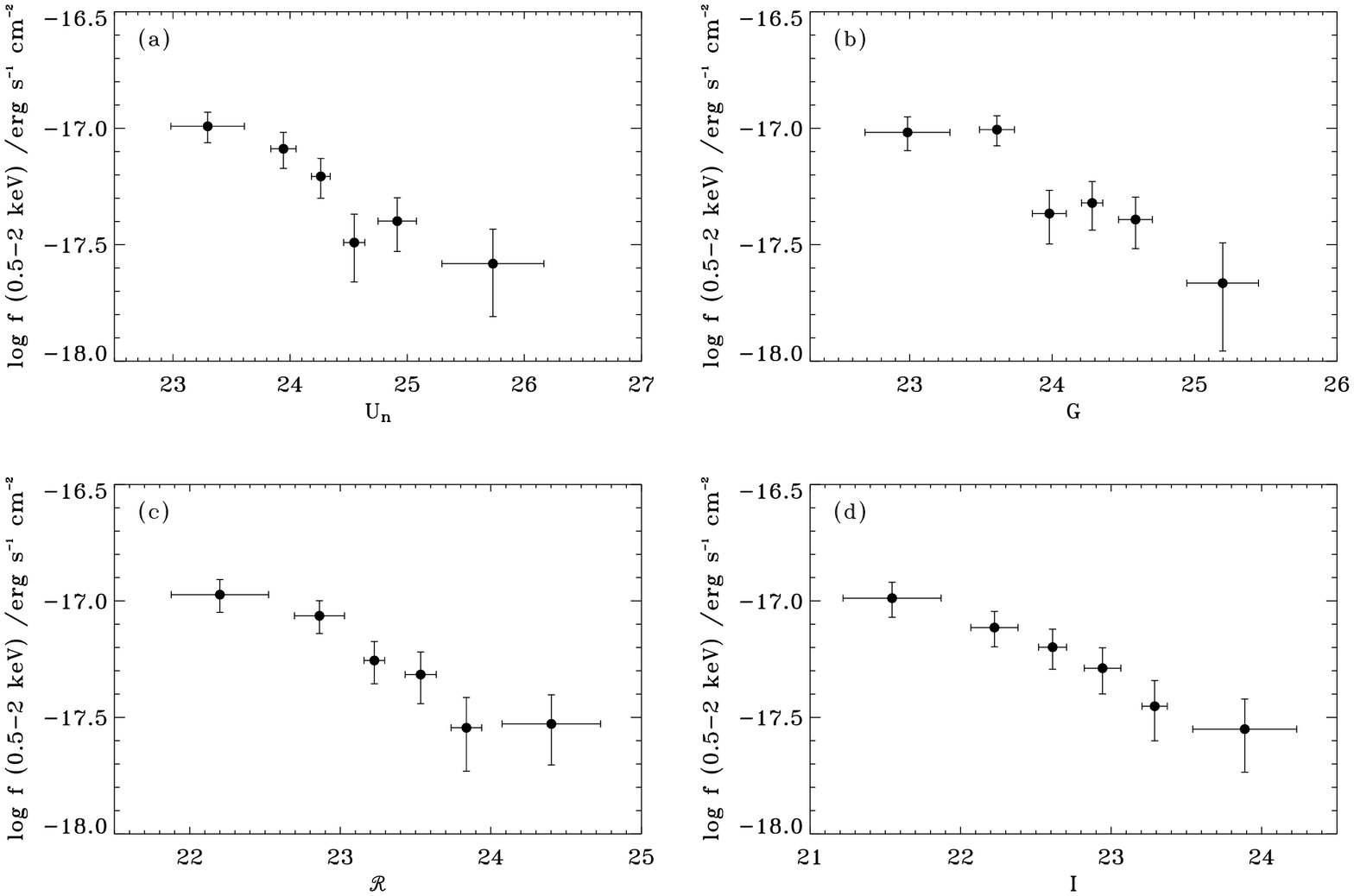}
\caption{ X-ray stacking results for undetected BBGs split into subsets according to broad
band magnitudes. 
Correlations of soft X-ray flux 
with (a)  $U_{\rmn{n}}$  magnitude, (b) $G$ magnitude, (c)
$\mathscr{R}$ magnitude, and (d) $I$ magnitude are shown.  
At $z$$\sim$1 $U_{\rmn{n}}, G, \mathscr{R}$ and $I$ filters correspond to rest-frame 
$\sim$1800~\AA,   $\sim$2400~\AA, $\sim$3400~\AA~and $\sim$4000~\AA~emission.
The $x$-axis errors are standard deviation of values in a given bin,
and the $y$-axis error bars are the Poisson errors from stacked soft band counts.}
\label{bbg_bins_mag}
\end{center}
\end{figure*}

The soft band signal for the undetected BBGs is strong enough to split
the galaxy sample into subsets according to broad band properties and
stack the X-ray emission for several bins, allowing the correlation of
X-ray emission with observed-frame optical magnitudes and colours to
be examined. In order to achieve reasonable signal-to-noise ratios,
for each broad band property we split the sample into 6 bins based on
$U_n, G, \mathscr{R}, I$ magnitude and $(U_n-G), (G-\mathscr{R}),
(\mathscr{R}-I)$ colour.  The range in redshift of the galaxies in our
sample, $\Delta z\simeq0.6$, means that a galaxies broad-band colour
is not just a function of its intrinsic properties but is also a
function of redshift. In order to investigate any correlations between
UV colour and X-ray emission without a contaminating redshift effect
we also calculate and bin according to the relative $(U_n-G), (G-\mathscr{R}),
(\mathscr{R}-I)$ colour excess for the galaxies.  For each colour
($U_n-G$ etc) the colour excess for each galaxy is calculated from the
difference between the galaxy colour and the median colour of all
galaxies with a redshift within $z=z\pm0.05$ of the galaxy
redshift. These values were binned and the mean colour excess for each
bin was shifted so that the lowest value bin, which presumably
contains the bluest and therefore least attenuated galaxies, was equal
to zero.
We denote this shifted colour excess by E(1800\AA--2400\AA),
E(2400\AA--3400\AA) and E(3400\AA--4050\AA) for the $(U_n-G),
(G-\mathscr{R})$ and $(\mathscr{R}-I)$ bands respectively, with the
wavelengths denoting the approximate rest-frame wavelength of the
emission in each filter. The stacking procedure used for the
magnitude, colour and colour excess bins was identical to that carried
out for the full samples of undetected sources.

The results of the subset stacking are shown in
Figures~4 (binning according to magnitude) and 5
(binning on colour). For each
quantity we show the exact mean 
0.5--2~keV flux per galaxy in a given bin, with errors,
regardless of whether or not the detection is considered significant. 
Low significance detections are easily identifiable by their
correspondingly large error bars.
The minimum flux for a 3$\sigma$ detection is 
$f_{0.5-2\rmn{keV}}\simeq3\times10^{-18}$~erg~s$^{-1}$~cm$^{-2}$ 
(log $f_{0.5-2\rmn{keV}}\simeq-17.5$).

We plot $U_{\rmn{n}}$, $G$, $\mathscr{R}$ and $I$
magnitude versus X-ray flux (Figure 4). 
Correlations between magnitude and soft X-ray
flux, in the sense that the brightest UV sources are also the
brightest X-ray sources, are seen for each filter. 
Spearman's rank and Pearson's linear correlation tests show each of
the relations to be highly significant (e.g. Pearson's
linear correlation coefficients were -0.93, -0.93, -0.95, -0.99, 
all with significance $>$99 per cent, for
$U_{\rmn{n}}$, $G$, $\mathscr{R}$ and $I$ respectively, Table 2). 
Rest-frame 1--4~keV X-ray emission and rest-frame  
UV emission are therefore related even before the 
effects of dust attenuation are applied. 
The results of the stacking in the six $U_{\rmn{n}}$ bins, which with
a rest-frame wavelength of $\sim$1800~\AA~at $z$$\sim$1 corresponds to
the wavelength commonly used for UV  measures of SFR, are shown in
Table 1. Only the weakest bin, 25.25 $<U_n\le$
27.03, does not have a significant ($>3\sigma$) X-ray detection.

The relations between X-ray flux and the colours of the BBGs is less clear.
For $U_{\rmn{n}}-G$, a measure of the spectral slope in the far UV,
the X-ray signal of the undetected sources is approximately flat 
for varying colour (Figure 5). 
With respect to $G-\mathscr{R}$ colour again the X-ray signal is
approximately flat, with perhaps some evidence
for a slight increase in X-ray flux with increasing redness. 
The results for $\mathscr{R}-I$, a measure of the spectral slope near 
the Balmer break, also show no evidence of a relation
with X-ray emission. Binning on colour excess, which removes any
redshift dependence in the colours, produced very similar
results to those shown in Figure 5: UV colour excess therefore also
showed no clear correlation with X-ray flux. Spearman's rank and
Pearson's linear correlation tests confirm the absence of a
significant correlation between colour and X-ray flux (Table 2).

\begin{figure*}
\begin{center}
\includegraphics[width=175mm]{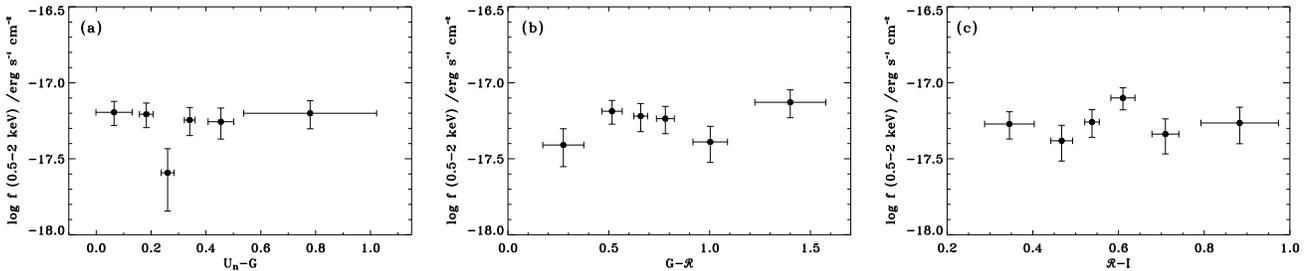}
\caption{ 
X-ray stacking results for undetected BBGs split into subsets according to broad
band colours. 
Correlations of soft X-ray flux 
with (a) $U_{\rmn{n}}-G$ colour, (b) $G-\mathscr{R}$
colour and (c) $\mathscr{R}-I$ colour are shown.
$U_{\rmn{n}}-G$ is a measure of the far-UV slope, $G-\mathscr{R}$ 
the near-UV slope and $\mathscr{R}-I$ the spectral slope near the
Balmer break. The  $x$-axis and $y$-axis error bars are as for Figure 4.} 
\label{bbg_bins_col}
\end{center}
\end{figure*}

\section{Discussion}

\subsection{Stacking signal and SFR estimate}

The X-ray emission from the undetected BBGs 
included in the stacking analysis should not be dominated by AGN 
as we excluded from the analysis all galaxies that were directly detected, 
equating to a L$_{\rmn{2-10~keV}} \gtrsim 10^{41}~\rmn{erg~s^{-1}}$ 
cut-off at $z$$\sim$1, therefore removing all moderate and luminous AGN. 
The mean stacking signal is certainly not dominated by a few AGN just below 
the detection threshold, as is demonstrated by the counts-in-cell distribution 
shown in Figure 3.
The presence of highly obscured AGN can also be ruled out from the
observed soft spectrum of the stacking signal. With significant stacked
detections in both the soft and hard bands we calculated
the mean hardness ratio (${\rmn{ HR = (H-S)/(H+S)}}$,  
where H and S are the net hard and soft band counts) of the sample. 
We find that HR=$-0.54\pm0.11$, which is consistent with a
$\Gamma\sim1.8\pm0.3$ unabsorbed power-law spectrum at $z$$\sim$1. 
While limited by large errors this result is consistent with the
observed spectra of local starburst and star forming galaxies
(e.g. \citealt{ptak99}) and is inconsistent with the hard spectrum expected if there
was a significant population of obscured AGN within the sample
(e.g. the spectrum of the X-ray background).   
This supports the assumption that the emission from the BBGs is a
result of star formation.
In saying this we should point out that the presence of low
luminosity AGN (LLAGN), with luminosity less than or comparable to the
host galaxies themselves, cannot be ruled out. Indeed given the
observed incidence of low level nuclear activity in the local Universe
\citep*{ho97} it is almost certain that there are more as yet
undiscovered AGN within the BBG sample. However the contribution of
such LLAGN to the mean BBG X-ray luminosity is expected to be small:
in a sample of nearby LLAGNs less than one third have X-ray emission
that is dominated by a compact nuclear source and the mean 2--10~keV
luminosity of the LLAGNs is less than $1$ per cent of that of the BBGs
\citep{ho01}. For the galaxy sample used here we therefore
assume that the observed emission is dominated by star formation
processes. For such a sample of galaxies the X-ray emission should 
be proportional to the SFR allowing the X-ray luminosity (normally 
2--10~keV) to be used as a SFR measure (N02; \citealt{ranalli03};
\citealt{grimm03}).
For situations such as 
the one here where only the total 2--10~keV luminosity of a galaxy 
or sample of galaxies is known, and the data render it impossible to 
resolve the HMXBs either spatially or spectrally, an X-ray--SFR 
relation with the form of the \citet{ranalli03} relation relating
the total 2--10~keV emission to SFR should be used. 
Only in local galaxies, where it is possible to resolve the HMXB 
population dominating the 2--10~keV flux, can the SFR be derived  
directly from the total luminosity emitted by HMXBs
(e.g. \citealt{grimm03}; \citealt{persic04}). 

An additional possible source of contamination is that of emission
from LMXBs which reflects the total stellar mass of a system as
opposed to the instantaneous SFR. For galaxies with low SFR emission
from LMXBs can dominated the flux and the X-rays will cease to
measure SFR \citep{grimm03}. 
While we do not attempt to model and separate out the contribution 
of LMXBs to our measured fluxes we can place approximate upper limits on the 
expected level of emission from LMXBs in our sample. 
Studies of the LMXB population in the Milky Way give an
estimate of the luminosity from LMXBs per solar mass of stars
\citep*{grimm02}. Using this result, and assuming that 
the \textit{mean} stellar mass of the BBGs is no greater than
$10^{11}$~M$_{\sun}$, gives an estimate of the maximum total
luminosity from LMXBs per BBG of $5\times10^{39}~\rmn{erg~s^{-1}}$.
Under these conditions the contribution of LMXBs to the 2--10~keV
luminosity would then be $\sim$15 per cent for the full sample, rising
to $\sim$30 per cent for the lowest significantly detected bins. We do
not therefore expect LMXBs to dominate the emission for this sample of
star forming galaxies. \textit{Spitzer} MIPS data, as well as ground
based K band imaging, can be used to probe the stellar masses of the BBGs
and will allow more stringent estimates of the LMXB contribution to be
made. This will be addressed in future work.

Given the expected low contribution from AGN and LMXBs to our 
detected signal we are confident that the X-ray emission from our sample
of undetected BBGs is indeed tracing star formation and we proceed according to 
this assumption. The observed X-ray fluxes and luminosities 
which correspond to rest-frame 1--4~keV emission therefore offer us a
measure of SFR at $z$$\sim$1 that is relatively unaffected by
attenuation by dust, for column densities of
N$_{\rmn{H}}<10^{22}~\rmn{cm}^{-2}$ which corresponds to A$_{\rmn{V}}\simeq50$.

Using the empirical relation between the total 2--10~keV luminosity 
and SFR found by \citet{ranalli03}, we find 
that the mean SFR per BBG is 
$6.0\pm0.6$~M$_{\sun}~\rmn{yr}^{-1}$ (Table 1),
consistent with that found by N02 for a smaller sample of BBGs using
the 1~Ms data. Comparison UV luminosities and estimates of the SFR are
calculated from the flux 
as measured in the  $U_{\rmn{n}}$ band, which at $z$$\sim$1 corresponds
to rest frame $\sim$1800~\AA, and the \citet{kennicutt98} UV
luminosity--SFR relation. 
These luminosities and SFRs have not
been corrected for dust attenuation and should therefore underestimate
the SFR compared to X-ray estimates, which better measure the
bolometric value. We find
$\rmn{SFR}_{\rmn{X}}/\rmn{SFR}_{\rmn{UV}}^{\rmn{uncor}}\sim$3 for the
full stacking sample, an estimate of the attenuation correction factor
for the UV.  This attenuation factor is slightly smaller than that
found by \citet{reddy04} for UV selected galaxies at $z$$\sim$2
($\rmn{SFR}_{\rmn{X}}/\rmn{SFR}_{\rmn{UV}}^{\rmn{uncor}}\sim$4.5) and
by \citet{N02} for $z$$\sim$3 LBGs
($\rmn{SFR}_{\rmn{X}}/\rmn{SFR}_{\rmn{UV}}^{\rmn{uncor}}\sim$5).
This difference is likely due to the mean SFRs of these samples 
being significantly larger than that considered here: the attenuation
factor is similar to the factor of 2.3 found for the 
SFR$<$20~M$_{\sun}~\rmn{yr}^{-1}$ subsample of \citet{reddy04}, which has a
more similar range in SFRs.

It is important to emphasize that the properties derived for our 
sample do not necessarily represent the BBGs as a whole. By 
excluding from our analysis all BBGs with significant direct X-ray 
detections we will undoubtedly have removed from the sample 
the galaxies with the highest SFRs, in addition to the AGN 
dominated sources. With a cut-off luminosity of 
$\sim$$10^{41}~\rmn{erg~s^{-1}}$ this is equivalent to removing all 
galaxies with SFR $\gtrsim20$~M$_{\sun}~\rmn{yr}^{-1}$. 
The mean SFR, and possibly also UV attenuation factor, of the BBGs as
a class is therefore likely to be higher than the values derived here 
(see \S4.2 for further discussion on observed correlations between SFR, 
attenuation and UV colour).   

\subsection{X-ray--UV correlations and implications}

\begin{figure}
\begin{center}
\includegraphics[width=75mm]{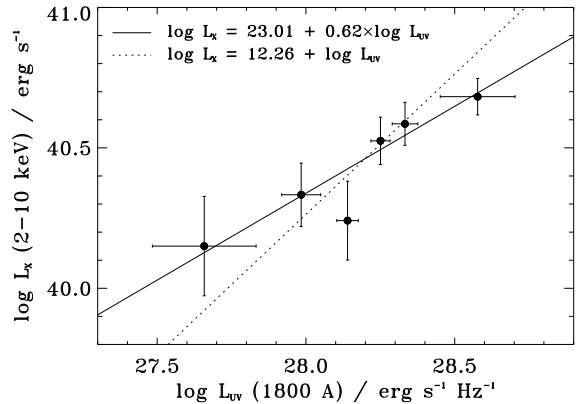}
\caption{2--10~keV X-ray luminosity vs. 1800~\AA~UV luminosity, for
binning on $U_{\rmn{n}}$ magnitude. Solid line
shows best-fitting power law  relation. Dotted line shows best-fitting power law
relation with 1-to-1 mapping between X-ray and UV luminosity.
}
\end{center}
\label{lum}
\end{figure}

\begin{figure*}
\begin{center}
\includegraphics[width=175mm]{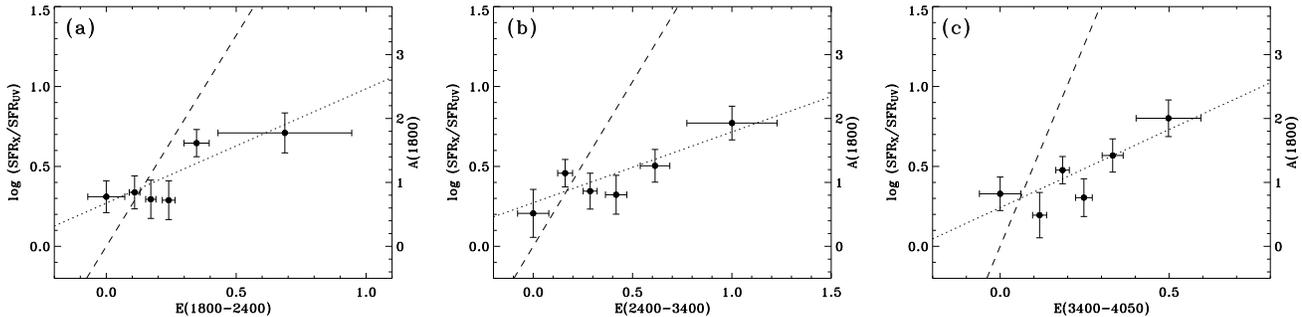}
\caption{Correlations between the ratio of X-ray derived SFR to UV derived 
SFR (and equivalently A$_{1800}$) and colour excess for
undetected BBGs. The UV SFR was calculated from the
$U_{\rmn{n}}$ band and is not corrected for attenuation. 
The sample was binned according to (a)
E(1800\AA--2400\AA), (b) E(2400\AA--3400\AA) and (c) E(3400\AA--4050\AA)
colour excess. Dotted lines show best-fitting relations with the form
log(SFR$_{\rmn{X}}$/SFR$_{\rmn{UV}}$) = a + (b$\times$colour excess).
Dashed lines show the \citet{cal00} attenuation law. 
}
\label{ratios}
\end{center}
\end{figure*}

The results of the subset stacking in \S3.3 
(Figure 4) clearly show that 
there is a correlation between X-ray flux and UV magnitudes.
These correlations, whereby the brightest UV sources are also the
brightest X-ray sources, hold over the rest-frame wavelength range
1800--4000~\AA. 
Good, approximately linear, correlations between X-ray luminosity and 
radio and far infrared (FIR) SFR indicators have been observed before for star
forming and starburst galaxies, indicating that the three wavelength regimes
are all following the SFR (\citealt{ranalli03}; \citealt{david92}). 
The correlation seen here between X-ray flux and 1800~\AA~emission also implies 
that both regimes are tracing the SFR.  
As discussed above, because rest-frame hard X-rays are relatively unaffected by
extinction they should reflect the intrinsic SFR without the need for
extinction corrections. On the face of it the observed correlation
with X-ray is therefore reassuring 
in terms of using 1800~\AA~to measure SFR at high redshift: in
general, for UV selected galaxies the sources with the apparent
highest SFR also appear to be the brightest in the UV. 
However given
that UV emission is very sensitive to dust attenuation the correlation
is a little surprising. Following the generally accepted picture of
higher SFR galaxies also being dustier (e.g. \citealt{adel00}) one
might expect the galaxies 
with large X-ray flux (representing high SFR) to have suppressed UV
flux. In Figure 6 we show the relation between X-ray and
1800~\AA~UV luminosity for the undetected galaxies. 
Spearman's rank and Pearson's linear correlation tests show a highly
significant correlation between log L$_{\rmn{X}}$ and log
L$_{\rmn{UV}}$ (Table 2). 
The best-fitting power-law is 
\begin{equation}
\rmn{log(L}_{\rmn{2-10~keV}}) = (0.62\pm0.16)\rmn{log(L}_{\rmn{UV}})+23.01\pm4.42 
\end{equation}
which has an acceptable chi-squared.
The best-fit to an exact linear slope 
\begin{equation}
\rmn{log(L}_{\rmn{2-10~keV}}) = \rmn{log(L}_{\rmn{UV}}) + 12.26 
\end{equation}
is also acceptable, at the 90 per cent level. 
With no evidence for increasing UV attenuation with increasing SFR, as
measured by the X-ray, the observed correlation between X-ray and UV is
counterintuitive to general picture of higher SFR galaxies being dustier.
The same effect is suggested with the 
$\rmn{SFR}_{\rmn{X}}/\rmn{SFR}_{\rmn{UV}}^{\rmn{uncor}}$ ratios in
Table 1: for increasing UV flux (and following on from
Figures~4 and 6, increasing X-ray flux and SFR) the mean UV attenuation
factors actually decrease (though within the errors the results are also
consistent with an approximately constant attenuation factor over the small range in 
SFRs covered here).
These results suggest that UV attenuation
may not necessarily be a direct function of SFR over the range of SFRs
considered here. 

The subset stacking also showed there to be no relation between
redness in the UV and X-ray flux (Figure 5) the 
immediate interpretation of which is that there is no direct
correlation between the colour of the UV selected galaxies and their
SFRs. Local starbursts exhibit a strong correlation between UV colour and
dust absorption, as measured for example by the ratio of far-infrared
to UV flux, implying that the starbursts redden as the dust absorption
increases and allowing UV attenuation to be estimated purely as a
function of UV colour (MHC99). Furthermore the locally derived
starburst attention laws appear to hold for high-$z$ UV-selected
galaxies such as LBGs (e.g. \citealt{seibert02}; \citealt{reddy04}). 
In order to investigate if such a correlation between colour and
attenuation is seen for the $z$$\sim$1 UV selected galaxies and if the
local relations hold we plot the ratio of X-ray to UV
derived SFR for subsets binned on colour excess (Figure 7). Since the X-ray
SFR estimates should be unaffected by absorption this ratio is a
measure of UV attenuation. In Figure 7 we also show the
corresponding attenuation coefficient, A$_{1800}$, related to
the SFR ratio by SFR$_{\rmn{X}}/$SFR$_{\rmn{UV}} =
10^{0.4\rmn{A}_{1800}}$. 
Correlations between
colour excess and attenuation are seen for each of the three bands,
with linear correlation tests (Table 2) revealing
the strongest correlation to be with E(2400\AA--3400\AA). For each, we fit a
linear function  log(SFR$_{\rmn{X}}$/SFR$_{\rmn{UV}}$) = a +
(b$\times$colour excess), where a and
b are constants, which is similar in form to that derived by
\citet{cal00}.  Each fit yields an acceptable $\chi^{2}$, with
E(2400\AA--3400\AA) producing the best fit. By binning the data on UV
colour excess and calculating the corresponding 
X-ray and UV SFRs for each bin we find the correlation between UV
colour, UV derived SFR and X-ray derived SFR to be  
\begin{equation}
\rmn{SFR_{X}} = (1.87^{+0.41}_{-0.61})\times10^{(0.44^{+0.20}_{-0.25})\rmn{E}(2400-3400)}\rmn{SFR_{UV}},
\end{equation}
where errors are for $\Delta\chi^2= 2.7$, or 90 per cent confidence,
for one interesting parameter.
For the SFRs covered in our sample, and over a range in colour excess of
$\Delta \rmn{E}\simeq1$, we find that the UV attenuation ranges from 0.5 to 2.0
magnitudes at 1800~\AA, corresponding to relatively modest UV
correction factors of $\sim1.5-6.5$. 
The observed colour--attenuation results, while certainly limited by
the number of fitting points and large errors, show clearly that UV
colour is correlated with dust attenuation for UV selected galaxies at
$z$$\sim$1. This key result allows a single UV colour to be used to
estimate UV attenuation corrections and attenuation corrected SFRs, as
shown above in Equation 3.

The attenuation coefficients, A$_{1800}$, and corresponding values of
log(SFR$_{\rmn{X}}$/SFR$_{\rmn{UV}}$) predicted by the reddening curve
and attenuation law of \citet{cal00} are
also shown in Figure 7. 
For a given reddening curve, the implied relationship between
attenuation, A$_{1800}$, and colour excess, E, is nearly independent of
the assumed underlying galaxy template. The Calzetti lines in Figure
7 are appropriate for galaxy spectral types Im, Sc, and Sb
which span the expected range for BBGs.
While the normalisation of our data points with
respect to absolute values of colour excess was approximate, and
therefore the relative normalisation between our data and the Calzetti
prediction is also approximate, the main conclusion is clearly
visible: the Calzetti law over predicts the attenuation corrections
compared to the observed values. Between E(1800\AA--2400\AA)=0 and
E(1800\AA--2400\AA)=1 approximately 7 magnitudes of extinction are predicted at
1800~\AA, compared to the observed difference of $\sim$2 magnitudes --
an over prediction of the range in UV luminosity and UV SFR correction
factors of $\sim$100. For a typical galaxy in our sample with
E(1800\AA--2400\AA)$=$0.25, the Calzetti law predicts A$_{1800}=$1.7 
compared to the observed A$_{1800}=$1.1 -- an over prediction in the UV SFR
correction factor of 1.6. 
For E(1800\AA--2400\AA), E(2400\AA--3400\AA)
and E(3400\AA--4050\AA) the probability that the data are derived from
the Calzetti model is $<0.01$ per cent, regardless of the exact
position of the Calzetti slope relative to the data points.  The
failure of the Calzetti attenuation law at predicting the attenuation
of the galaxies in our sample is most likely a result of the inherent
differences in the sample of galaxies used here and those from which
the attenuation laws were calculated.  The attenuation laws were
derived for local starburst galaxies for which the UV colours are
predominantly affected by dust content and obscuration. BBG selection
on the other hand identifies Sb- and Sc-type star forming galaxies as
well as starbursts \citep{adel04} and therefore includes galaxies with
more quiescent star formation, and a range of star formation histories
and stellar population ages which affect the UV colours as much as the
dust content. Such samples of galaxies clearly do not follow the
starburst attenuation laws.

\section{Summary}
With the 2~Ms observation of the HDF-N region we have examined the X-ray
properties of a large sample of UV-selected, star forming galaxies with
$0.74<z<1.32$. In this paper we dealt exclusively with the 216 galaxies
in the sample that are not directly detected in X-ray and whose 
emission should be predominately due to star formation processes
(analysis of the detected BBGs will be presented in an upcoming 
paper, Laird et al., in preparation). Here we summarise our main
results:
\begin{enumerate}
\item Stacking the soft band flux from the undetected BBGs resulted in
a highly significant, 14.9$\sigma$ detection. The detected signal
corresponds to a mean flux per galaxy  
of $5.51\pm0.49\times10^{-18}$~erg~s$^{-1}$~cm$^{-2}$ (0.5--2~keV) and
a mean  luminosity of 
L$_{2-10 \rmn{keV}} = 2.97\pm0.26\times10^{40}$~erg~s$^{-1}$.

\item Stacking the hard band flux also resulted in a significant
detection of low level emission. We find a mean 2-10~keV flux per
undetected BBG of 
$8.87\pm2.91\times10^{-18}$~erg~s$^{-1}$~cm$^{-2}$. 

\item The hardness ratio of the sample, HR$=-0.54\pm0.11$, is consistent
with a $\Gamma\sim1.8\pm0.3$ unabsorbed power-law spectrum. This is
consistent with the integrated spectrum from HMXBs and
supports the general interpretation that the detected emission is due
to normal star formation processes. 

\item The mean X-ray derived SFR of the sample of undetected BBGs is
$6.0\pm0.6$ M$_{\sun}~\rmn{yr}^{-1}$. The ratio of X-ray derived SFR
to UV derived SFR, an indication of the average UV attenuation factor,
is 3. 

\item Splitting the sample into subsets based on observed-frame optical
magnitudes and colours and stacking the X-ray flux in each bin
produced significant detections, 
allowing correlations between X-ray and UV emission to be
examined. We find clear correlations between X-ray flux and
rest-frame UV flux (over the range $\sim$1800--4000~\AA) such that
galaxies that are brighter in the UV are also brighter in X-ray. There
is no evidence for a correlation between UV colour or UV colour excess
and X-ray flux. 

\item Correlations between colour excess and the ratio of
X-ray-to-UV SFRs (which measures UV attenuation) are seen, whereby the
ratio of X-ray-to-UV increases with UV redness. Using the data we 
derive a relation for estimating UV attenuation corrections from
UV colour excess.  

\item The observed relation between colour excess and attenuation is
found to be inconsistent with the \citet{cal00} and MHC99 extinction
curve which over predict the range in attenuation corrections for the
galaxies in this sample by a factor of $\sim$100. 

\end{enumerate}

\section*{Acknowledgments}
We thank the referee for helpful comments that improved the
manuscript and the \textit{Chandra} X-ray Center for assistance
with some analysis and software issues. D. Rosa-Gonz{\' a}lez is
thanked for many useful discussions. ESL acknowledges the support of a
PPARC Studentship and a University of London Valerie Myerscough Prize.

\label{lastpage}


\begin{thebibliography}{99}

\bibitem[\protect\citeauthoryear{Adelberger \& 
Steidel}{2000}]{adel00} Adelberger K.~L., Steidel C.~C., 2000, 
ApJ, 544, 2
\bibitem[\protect\citeauthoryear{Adelberger et al.}{2004}]{adel04}
Adelberger K.~L., Steidel C.~C., Shapley A.~E., Hunt M.~P., Erb D.~K.,
Reddy N.~A., Pettini M., 2004, ApJ, 607, 266
\bibitem[\protect\citeauthoryear{Alexander et al.}{2003}]{A03} Alexander D.~M.~et 
al. 2003, AJ, 126, 539 
\bibitem[\protect\citeauthoryear{Bell}{2002}]{bell02} Bell 
E.~F., 2002, ApJ, 577, 150 
\bibitem[\protect\citeauthoryear{Brandt et al.}{2001}]{brandt01} Brandt W.~N., 
Hornschemeier  A.~E., Schneider  D.~P., Alexander  D.~M., Bauer  F.~E., 
Garmire  G.~P., Vignali  C., 2001, ApJL, 558, L5 
\bibitem[\protect\citeauthoryear{Calzetti, Kinney, \& 
Storchi-Bergmann}{Calzetti et al.}{1994}]{cal94} Calzetti D., Kinney A.~L., 
Storchi-Bergmann T., 1994, ApJ, 429, 582 
\bibitem[\protect\citeauthoryear{Calzetti et 
al.}{2000}]{cal00} Calzetti D., Armus L., Bohlin R.~C., 
Kinney A.~L., Koornneef J., Storchi-Bergmann T., 2000, ApJ, 533, 682 
\bibitem[\protect\citeauthoryear{David, Jones, \& 
Forman}{David et al.}{1992}]{david92} David L.~P., Jones C., Forman W., 1992, 
ApJ, 388, 82 
\bibitem[\protect\citeauthoryear{Fabbiano}{1989}]{fabbiano89} 
Fabbiano G., 1989, ARA\&A, 27, 87 
\bibitem[\protect\citeauthoryear{Giavalisco}{2002}]{giavalisco02}
Giavalisco M., 2002, ARA\&A, 40, 579 
\bibitem[\protect\citeauthoryear{Grimm, Gilfanov, \& 
Sunyaev}{Grimm et al.}{2002}]{grimm02} Grimm H.-J., Gilfanov M., Sunyaev R., 
2002, A\&A, 391, 923 
\bibitem[\protect\citeauthoryear{Grimm, Gilfanov, \& 
Sunyaev}{Grimm et al.}{2003}]{grimm03} Grimm H.-J., Gilfanov M., Sunyaev R., 
2003, MNRAS, 339, 793 
\bibitem[\protect\citeauthoryear{Ho, Filippenko, \& 
Sargent}{Ho et al.}{1997}]{ho97} Ho L.~C., Filippenko A.~V., Sargent 
W.~L.~W., 1997, ApJ, 487, 568 
\bibitem[\protect\citeauthoryear{Ho et al.}{2001}]{ho01} Ho 
L.~C., et al., 2001, ApJ, 549, L5
\bibitem[\protect\citeauthoryear{Hornschemeier et 
al.}{2001}]{horn01} Hornschemeier A.~E., et al., 2001, ApJ, 
554, 742 
\bibitem[\protect\citeauthoryear{Hughes et al.}{1998}]{hughes98} 
Hughes D.~H., et al., 1998, Nat, 394, 241 
\bibitem[\protect\citeauthoryear{Kennicutt}{1998}]{kennicutt98} 
Kennicutt R.~C., 1998, ARA\&A, 36, 189 
\bibitem[\protect\citeauthoryear{Lehmer et al.}{2005}]{lehmer05} 
Lehmer B.~D., et al., 2005, AJ, 129, 1 
\bibitem[\protect\citeauthoryear{Lilly et al.}{1996}]{lilly96} 
Lilly S.~J., Le Fevre O., Hammer F., Crampton D., 1996, ApJ, 460, L1 
\bibitem[\protect\citeauthoryear{Madau et al.}{1996}]{madau96} 
Madau P., Ferguson H.~C., Dickinson M.~E., Giavalisco M., Steidel C.~C., 
Fruchter A., 1996, MNRAS, 283, 1388 
\bibitem[\protect\citeauthoryear{Marshall et 
al.}{2004}]{marshall04} Marshall H.~L., Tennant A., Grant C.~E., 
Hitchcock A.~P., O'Dell S.~L., Plucinsky P.~P., 2004, SPIE, 5165, 497 
\bibitem[\protect\citeauthoryear{Meurer, Heckman, \& 
Calzetti}{Meurer et al.}{1999}]{meurer99} Meurer G.~R., Heckman T.~M., Calzetti 
D., 1999, ApJ, 521, 64
\bibitem[\protect\citeauthoryear{Nandra et al.}{2002}]{N02} Nandra
K., Mushotzky  R.~F., Arnaud  K., Steidel  C.~C., Adelberger  K.~L.,
Gardner  J.~P., Teplitz  H.~I., Windhorst  R.~A., 2002, ApJ, 576, 625 
\bibitem[\protect\citeauthoryear{Nandra et al.}{2005}]{nandra05} 
Nandra K., et al., 2005, MNRAS, 356, 568 
\bibitem[\protect\citeauthoryear{Persic et al.}{2004}]{persic04}
Persic M., Rephaeli Y., Braito V., Cappi M., Della Ceca R., Franceschini 
A., Gruber D.~E., 2004, A\&A, 419, 849 
\bibitem[\protect\citeauthoryear{Ptak et al.}{1999}]{ptak99} 
Ptak A., Serlemitsos P., Yaqoob T., Mushotzky R., 1999, ApJS, 120, 179 
\bibitem[\protect\citeauthoryear{Ranalli, Comastri, \& 
Setti}{Ranalli et al.}{2003}]{ranalli03} Ranalli P., Comastri A., Setti G., 2003, 
A\&A, 399, 39 
\bibitem[\protect\citeauthoryear{Reddy \& 
Steidel}{2004}]{reddy04} Reddy N.~A., Steidel C.~C., 2004, ApJ, 
603, L13
\bibitem[\protect\citeauthoryear{Seibert, Heckman, \& 
Meurer}{Seibert et al.}{2002}]{seibert02} Seibert M., Heckman T.~M., Meurer 
G.~R., 2002, AJ, 124, 46 
\bibitem[\protect\citeauthoryear{Stark et al.}{1992}]{stark92} Stark
A.~A., Gammie C.~F., Wilson  R.~W., Bally  J., Linke  R.~A., Heiles
C., Hurwitz  M., 1992, ApJS, 79, 77 
\bibitem[\protect\citeauthoryear{Steidel et 
al.}{1996}]{steidel96} Steidel C.~C., Giavalisco M., Pettini M., 
Dickinson M., Adelberger K.~L., 1996, ApJ, 462, L17 
\bibitem[\protect\citeauthoryear{Steidel et 
al.}{1999}]{steidel99} Steidel C.~C., Adelberger K.~L., 
Giavalisco M., Dickinson M., Pettini M., 1999, ApJ, 519, 1 
\bibitem[\protect\citeauthoryear{Steidel et al.}{2003}]{steidel03} Steidel  C.~C., 
Adelberger  K.~L., Shapley  A.~E., Pettini  M., Dickinson  M.,
Giavalisco  M., 2003, ApJ, 592, 728
\bibitem[\protect\citeauthoryear{Steidel et 
al.}{2004}]{steidel04} Steidel C.~C., Shapley A.~E., Pettini M., 
Adelberger K.~L., Erb D.~K., Reddy N.~A., Hunt M.~P., 2004, ApJ, 604, 534 
\bibitem[\protect\citeauthoryear{Wirth et al.}{2004}]{wirth04} 
Wirth G.~D., et al., 2004, AJ, 127, 3121 

\end{thebibliography}
\end{document}